%% file: conference_101719.tex
\documentclass[sigconf]{acmart}
\copyrightyear{2024}
\acmYear{2024}
\setcopyright{rightsretained}
\acmConference[ICSE-Companion '24]{2024 IEEE/ACM 46th International
Conference on Software Engineering: Companion Proceedings}{April 14--20,
2024}{Lisbon, Portugal}
\acmBooktitle{2024 IEEE/ACM 46th International Conference on Software
Engineering: Companion Proceedings (ICSE-Companion '24), April 14--20, 2024,
Lisbon, Portugal}\acmDOI{10.1145/3639478.3640023}
\acmISBN{979-8-4007-0502-1/24/04}

\usepackage[T1]{fontenc}

\usepackage{hyperref}
\usepackage{algorithmic}
\usepackage{graphicx}
\usepackage{textcomp}
\usepackage{xcolor}
\usepackage{xspace}
\usepackage{paralist}

\usepackage{color, colortbl}
\definecolor{Gray}{gray}{0.9}

\def\BibTeX{{\rm B\kern-.05em{\sc i\kern-.025em b}\kern-.08em
    T\kern-.1667em\lower.7ex\hbox{E}\kern-.125emX}}

\newcommand{\TODO}[1]{\textcolor{red}{#1}\GenericWarning{}{LaTeX Warning: TODO: #1}}\newcommand\todo\TODO

\newcommand{\name}{\textsc{GitBug-Actions}\xspace}

\begin{document}

\title{GitBug-Actions: Building Reproducible Bug-Fix Benchmarks with GitHub Actions}

\author{Nuno Saavedra$^*$}
\affiliation{
    \institution{INESC-ID/IST, University of Lisbon}
    \city{Lisbon}
    \country{Portugal}
}
\email{nuno.saavedra@tecnico.ulisboa.pt}

\author{André Silva$^*$}
\affiliation{
    \institution{KTH Royal Institute of Technology}
    \city{Stockholm}
    \country{Sweden}
}
\email{andreans@kth.se}

\author{Martin Monperrus}
\affiliation{
    \institution{KTH Royal Institute of Technology}
    \city{Stockholm}
    \country{Sweden}
}
\email{monperrus@kth.se}

\begin{abstract}
Bug-fix benchmarks are fundamental in advancing various sub-fields of software engineering such as automatic program repair (APR) and fault localization (FL).
A good benchmark must include recent examples that accurately reflect technologies and development practices of today.
To be executable in the long term, a benchmark must feature test suites that do not degrade overtime due to, for example, dependencies that are no longer available. 
Existing benchmarks fail in meeting both criteria.
For instance, Defects4J, one of the foremost Java benchmarks, last received an update in 2020.
Moreover, full-reproducibility has been neglected by the majority of existing benchmarks.
In this paper, we present \name: a novel tool for building bug-fix benchmarks with modern and fully-reproducible bug-fixes.
\name relies on the most popular CI platform, GitHub Actions, to detect bug-fixes and smartly locally execute the CI pipeline in a controlled and reproducible environment.
To the best of our knowledge, we are the first to rely on GitHub Actions to collect bug-fixes.
To demonstrate our toolchain, we deploy \name to build a proof-of-concept Go bug-fix benchmark containing executable, fully-reproducible bug-fixes from different repositories.
A video demonstrating \name is available at: \url{https://youtu.be/aBWwa1sJYBs}.
\end{abstract}


\maketitle

\def\thefootnote{*}\footnotetext{These authors contributed equally to this work.}\def\thefootnote{\arabic{footnote}}

\section{Introduction}
\label{sec:intro}

\input{introduction}

\section{The GitBug-Actions Workflow}

\input{defects4all}

\section{GitBug-Actions on the Go}

\input{evaluation}

\section{Related Work}

\input{related_work}

\section{Conclusion}

We present \name, a tool for building bug-fix benchmarks based on GitHub Actions.
\name builds fully-reproducible bug-fix benchmarks by using GitHub Actions to both collect recent examples and create fully-reproducible environments for the collected bug-fixes.
To the best of our knowledge, we are the first to leverage GitHub Actions to build bug-fix benchmarks.

We demonstrate \name's effectiveness by building a benchmark of Go bug-fixes from January 2023.
In total, \name collects 21 fully-reproducible bug-fixes from 13 different repositories on GitHub.
Future work could extend \name by applying automatic patch minimization to the collected patches. 
\name is made publicly available at \url{https://github.com/gitbugactions/gitbugactions}.

\section*{Acknowledgements}
This work was partially supported by the Wallenberg AI, Autono\-mous Systems and Software Program (WASP) funded by the Knut and Alice Wallenberg Foundation.
This work was supported by national funds through FCT, Fundação para a Ciência e a Tecnologia, under grant BD/04736/2023.
This work was supported by national funds through FCT, Fundação para a Ciência e a Tecnologia, under project UIDB/50021/2020 (DOI:10.54499/UIDB/50021/2020).
A special thanks to Bernardo Conde for the help in debugging Linux issues.

\bibliographystyle{IEEEtran}
\bibliography{IEEEabrv,references}

\end{document}

%% file: introduction.tex
Bug-fix benchmarks play a pivotal role in advancing the field of software engineering by providing essential resources for evaluating methodologies in various sub-fields, such as automatic program repair (APR) and fault localization (FL)~\cite{sim2003using, wright2010validity}.
A bug-fix is a software modification that fixes an existing defect, aligning the program's behavior with the intended specification.
It is represented by a pair of commits: a buggy commit and a subsequent fixing commit. 
For example, Defects4J~\cite{just2014defects4j} is a widely adopted bug-fix benchmark that has greatly served software engineering research in the past decade.
Good benchmarks must be representative and rigorous.

First, benchmarks must ensure that the research community studies relevant problems of today in modern software.
Rather than relying on outdated examples, benchmarks must include bug-fixes that are reflective of modern development practices and that use modern programming languages and build tools. 
Moreover, benchmarks of recent bugs help reduce the risk of data leakage, that is evaluating large language models (LLMs) techniques with data seen at training time~\cite{jacovi2023stop, zhang2023critical}.

Second, reproducible benchmarks ensure that such studies can be validated by third-parties today but also in the future.
While reproducibility is fundamental in the scientific method, bug-fix benchmarks have failed to retain it over time.
For example, Zhu and Rubio-González~\cite{zhu2023reproducibility} show that reproducibility in bug-fix benchmarks varies between 26.6\% and 96.9\%, with none achieving full-reproducibility.

Continuous Integration (CI) systems have served as a valuable source of bug instances~\cite{tomassi2019bugswarm, madeiral2019bears}.
By automating the build and testing processes, CI systems precisely capture developer-specified environments from which bug-fix samples can be collected and reproduced.
GitHub Actions is the most popular CI system~\cite{JetBrainsCI}.
It offers GitHub users an integrated platform for defining CI workflows to build and execute test suites.
Our key insight is that GitHub Actions is a valuable resource for creating high-quality bug-fix benchmarks.

In this paper, we propose \name a novel methodology for building bug-fix benchmarks based on GitHub Actions.
\name relies on GitHub Actions to identify and run bug-fixes in the same environment as the one defined by developers.
By using GitHub Actions, \name collects recent bug-fixes, that reflect the variety of real-world cases due to the CI system's widespread adoption.

Moreover, \name preserves the collected bug-fixes in fully-reproducible formats.
This is achieved by locally executing the bug-fixes in the environment specified by the developers in the GitHub Actions workflow, and then storing the necessary files, in particular all software dependencies, to re-execute the bug-fixes offline in the same environment. 
In this way, \name builds bug-fix benchmarks that uphold scientific standards w.r.t. reproducibility.
The bug-fixes in \name are designed to be executable for eternity.

To validate the concept \name, we build a proof-of-concept benchmark of Go bug-fixes from January 2023.
In total, \name successfully collects bug-fixes that are 1) executable, 2) fully-reproducible and 3) come from different repositories.

To summarize, our contributions are:
\begin{itemize}
    \item 
        An original workflow for building bug-fix benchmarks using GitHub Actions, called \name.
        To the best of our knowledge, we are the first to use GitHub Actions to build bug-fix benchmarks.

    \item 
        \name's implementation, made publicly available for researchers to build benchmarks in their programming language and stack of choice: \url{https://github.com/gitbugactions/gitbugactions}.

    \item A proof-of-concept benchmark of Go bug-fixes from January 2023, collected by \name.
    The benchmark contains 21 fully-reproducible bug-fix commits from 13 different repositories. 
\end{itemize}

\begin{figure*}[t!]
    \centering
    \includegraphics[width=\textwidth]{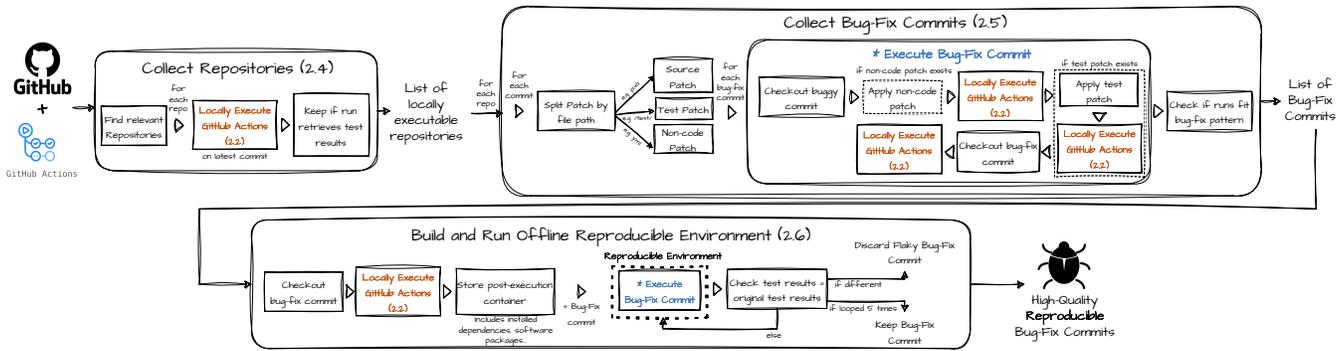}
    \caption{Overview of \name, a novel methodology to collect bug-fixes based on Github Actions.}
    \label{fig:overview}
\end{figure*}

%% file: defects4all.tex
In this section, we present the design of \name, a novel tool for collecting fully-reproducible bug benchmarks.

\subsection{Overall Workflow}

\name builds on top of the ability to locally execute GitHub Actions (Section \ref{sec:localrepro}) for multiple programming languages and build systems (Section \ref{sec:tailoring}).
Its pipeline, shown in Figure \ref{fig:overview}, is composed of three stages:
\begin{inparaenum}
    \item \textit{Collect Repositories} (Section \ref{sec:collectrepos})
    \item \textit{Collect Bug-Fix Commits} (Section \ref{sec:collectbugfix})
    \item \textit{Build and Run Offline Reproducible Environment} (Section \ref{sec:filterbugs}).
\end{inparaenum}
The final produce of \name is a benchmark containing high-quality reproducible bug-fix commits.

\subsection{Locally Execute GitHub Actions}
\label{sec:localrepro}

GitHub Actions is a CI service provided by GitHub, the most popular according to a 2022 survey~\cite{JetBrainsCI} by JetBrains.
GitHub Actions' builds are declared by \textit{workflows}.
Workflows are configurable YAML documents that define:
\begin{inparaenum}
    \item the events that trigger its run (e.g., a \textit{git push} event);
    \item the jobs that will run;
    \item the environment in which the jobs will run, typically a VM in Azure;
    and
    \item the list of steps to be run in each job, which can either run shell commands or a reusable third-party action.
\end{inparaenum}

Executing bug-fixes is a fundamental aspect for a bug-fix benchmark, since one needs to execute test suites to verify the correctness or incorrectness of a program.
To build a fully-reproducible benchmark, execution needs to be local to not depend on third-party services.
To locally execute GitHub Actions, \name relies on the well-established open-source tool Act~\footnote{\url{https://github.com/nektos/act}}.
Act uses docker images that imitate GitHub's proprietary execution environments.
For each build, it initializes containers based on these images and the environment setup defined in the workflow to run.

\name builds on top of Act.
First, \name identifies the workflows which contain test execution commands (e.g. \textsc{maven test}) by static analysis of the YAML build file.
Then, the identified workflows are modified as follows:
the operating system is set to \textit{ubuntu-latest} to ensure compatibility with Act;
matrix configurations are simplified to the first existing configuration;
only jobs containing test commands and their dependencies (i.e. as stated by the \textit{needs} operator) are kept;
test commands are instrumented to ensure the generation of test reports.
After executing Act on the instrumented workflows, \name parses the test reports and returns the test execution result.
Test execution results are useful in verifying whether the program respects the expected behavior defined by the test suite.

\subsection{Tailoring Data Collection per Programming Language and Build System}
\label{sec:tailoring}

\name is designed to be programming language and build system agnostic.
It builds a unified abstraction layer that standardizes the interaction with test execution workflows.
Such layer overcomes the variability associated with such diversity.
For each build tool, the necessary information to support a given programming language and build system is:
\begin{inparaenum}
    \item how to distinguish source-code files from test files,
    \item how to identify a test execution command, and
    \item how to retrieve test execution results.
\end{inparaenum}

\name already supports Java (Maven and Gradle), Python (pytest and unittest) and Go.

\subsection{Collect Repositories}
\label{sec:collectrepos}

\name employs a systematic approach to identify locally reproducible open-source repositories on GitHub.

First, it selects repositories that meet specific criteria per a GitHub search query\footnote{\url{https://docs.github.com/en/search-github/searching-on-github/searching-for-repositories}}, for example, the number of stars a repository has received from users.
Second, it attempts to execute the repositories' GitHub Actions as described in Section \ref{sec:localrepro}.
Repositories are retained if and only if \name can retrieve test execution results from the executed workflows.
Successful execution and retrieval of test reports indicate that the executed actions are executing tests, rendering the repository as a potential source of bug instances for further analysis.
We recall that test execution results are fundamental for building a bug-fix benchmark since they are used to verify the correctness or incorrectness of a program.

\subsection{Collect Bug-Fix Commits}
\label{sec:collectbugfix}

\name searches the commit history of each of the locally executable repositories for bug-fix commits.
A bug-fix is a pair of commits (${commit}_{t-1}$, ${commit}_{t}$) such that ${commit}_{t-1}$ corresponds to the buggy version and ${commit}_{t}$ corresponds to the fixed version.
\name collects behavioral bug-fixes, so buggy programs must have at least one failing test case, and fixed programs must have a passing test suite.

To identify bug-fix commits, \name first splits each candidate commit's patch into three patches by analyzing the path and extension of each modified file:
\begin{inparaenum}
    \item \textit{Source Patch}: Contains changes in the source code under test,
    \item \textit{Test Patch}: Contains changes in test cases, and
    \item \textit{Non-Code Patch}: Contains changes in non-source files such as documentation files and configuration files.
\end{inparaenum}
This split is important for two reasons.
First, developers often couple non-source code changes in commits that fix behavioral bugs.
As a result, bug-fix commits can become polluted with changes that are not relevant to the program's behavior.
By isolating these changes, \name collects higher-quality bug-fix patches.
Second, when fixing a bug, developers often introduce test changes to validate them.
Isolating test changes is thus crucial in building a buggy version that has failing test cases.

In detail, \name matches candidate bug-fix commit pairs with two patterns based on their characteristics and test execution results.
These patterns are based on the Bears benchmark~\cite{madeiral2019bears}:
\begin{inparaenum}
    \item \textbf{Passing Commit + Passing Commit with Source Changes and Test Changes}:
    In this scenario, the human-developer introduces bug-fix changes with ${commit}_{t}$, as well as test changes to validate them.
    We look for pairs of subsequent commits (${commit}_{t-1}$, ${commit}_{t}$) which have passing CI builds.
    ${commit}_{t}$ must introduce both source code and test code changes, which must not be removal only.
    ${commit}_{t-1}$'s build must fail only when the test changes from ${commit}_{t}$ are applied.

    \item \textbf{Failing Commit + Passing Commit with Source Changes}:
    In this scenario, the human-developer introduces bug-fix changes such that the program adheres to the entire pre-existing specification.
    We look for pairs of commits (${commit}_{t-1}$, ${commit}_{t}$) with the following CI status:
    ${commit}_{t-1}$ has a failing CI build and
    ${commit}_{t}$ has a passing CI build.
    ${commit}_{t}$ must introduce source code changes and not change the test suite.

\end{inparaenum}

\subsection{Build Offline Reproduction Environments}
\label{sec:filterbugs}

A big challenge in collecting bug benchmarks lies in ensuring all bugs remain reproducible in the future~\cite{zhu2023reproducibility}.
Due to the prevalence of third-party code in complex applications, typically retrieved during build time, reproducibility often depends on outside actors (e.g. package maintainers and repositories) which may become unpredictably unavailable.
Another issue lies in flaky tests, which introduce non-determinism in experimental reproductions.

\name builds offline reproduction environments to safeguard reproducibility as a key component of the collected benchmarks.
This is achieved in two steps.
First, \name stores the docker container's state after locally executing the tests workflow on ${commit}_{t}$.
The stored container's state includes all installed software packages required to run the same test workflow again without access to the internet for both the buggy and fixed versions, due to the normalization of non-code changes which include any changes in dependencies.
Being able to reproduce test execution offline is essential for reproducibility since required software packages might become unavailable in the future.
Second, \name executes each collected bug-fix commits $K$ times.
Only those bugs that yield the exact same test results across all $K$ executions are kept in the benchmark, effectively removing cases with flaky tests that introduce non-determinism in the benchmark.

%% file: evaluation.tex
We instantiate \name to create a benchmark containing bug-fix commits from January 2023 written in the programming language Go.
Due to \name's workflow abstraction, the required logic to handle Go workflows and extract test execution results is implemented in a single file\footnote{\url{https://github.com/gitbugactions/gitbugactions/blob/e49ff5bdf57d08dab29d3601a3abe942e04a5dbd/gitbugactions/actions/go/go_workflow.py}}.
The Go benchmark is available on Zenodo: \url{https://zenodo.org/records/10034611}. An interactive visualization of the benchmark is available at \url{https://nfsaavedra.github.io/gba-on-the-go}.

We run the entire \name's pipeline.
In the \textit{Collect Repositories} step, the following criteria are used to select repositories from GitHub:
\begin{inparaenum}
    \item Programming Language: Main programming language must be Go per the GitHub metadata.
    \item Popularity: At least 50 stars. The star count serves as an indicator of popularity and community engagement, ensuring that selected repositories have a certain level of relevance and activity.
    \item Size: Repository is less than 200MB. This criterion serves for storage space efficiency.
\end{inparaenum}
\textit{Collect Bug-Fix Commits} is configured to only consider commits from January 2023.
\name is deployed with 32 parallel workers on a machine with an AMD EPYC 7742 64-Core Processor and 512GB of memory.
The pipeline's run starts on October 10th 2023 and takes approximately 54 hours to complete.
Estimating a full-year benchmark collection equates to 516 hours or 21.5 days, considering the one-off cost of the \textit{Collect Repositories} step is 12 hours.
We believe this cost is reasonable for collecting a full-year benchmark.

\emph{Collect Repositories. } 
\name finds 21,891 GitHub repositories in Go that follow the aforementioned criteria.
For these, if there exists GitHub Actions that execute tests on the latest commit of the default branch, \name attempts to locally execute the Go test suite.
\name locally executes the test workflow of 3,465 repositories that have a single test GitHub Action.
In total, \name is able to execute and obtain test execution results of 1,626/3,465 (46.9\%) repositories.

\emph{Collect Bug-Fix Commits.}
For each of the 1,626 repositories,
\name finds 9,567 commits from January 2023.
Recall that, for each commit, \name locally executes the associated test workflow.
A pair of subsequent commits is considered a bug-fix if they match one of the bug-fix patterns explained in Section \ref{sec:collectbugfix}.
In total, \textit{Collect Bug-Fix Commits} identifies 33 bug-fix commits from 20 different repositories per the considered patterns.

\emph{Build Offline Reproduction Environments.}
This step builds and runs an offline reproducible environment for each of the 33 identified bug-fix commits.
Recall that, for a bug-fix to be included in the benchmark, it must be fully-reproducible, meaning it must run in offline isolation and not have flaky tests $K$ times.
We set $K$ to 5.
This rules out 6 commits that have flaky tests and 6 commits that were not able to run in offline isolation.
We build the fully-reproducible images that contain all dependencies.
In total, \textit{Build Offline Reproduction Envrionments} finds 21 fully-reproducible bug-fixes from 13 different repositories.
To validate the experimental benchmark, the two first co-authors manually verified each bug-fix. Out of the 21, 17 are considered to be high-quality and would be included in a real benchmark.
Researchers could run the system over a timespan of several years to the amount of bug-fixes they need.
This proof-of-concept benchmark demonstrates that \name can be configured for any programming language with little effort.

%% file: related_work.tex
Several bug-fix benchmarks suitable for different purposes and with a diverse range of properties have been proposed in the literature.
QuixBugs~\cite{lin2017quixbugs}, 
Codeflaws~\cite{tan2017codeflaws}, 
Code4Bench~\cite{majd2019code4bench}, 
RunBugRun~\cite{prenner2023runbugrun}, and 
EvalGPTFix~\cite{zhang2023critical} 
are benchmarks constructed from coding competition websites.
Such problems, while real, are not representative of those that developers face in complex software systems.
Others, like HumanEval-Java~\cite{jiang2023impact}, rely on artificially injected bugs which are, by nature, not real.

Benchmarks such as FixJS~\cite{csuvik2022fixjs} and Minecraft~\cite{avulaminecraft} do not include test suites for each bug-fix instance, rendering them unsuitable for studies reliant on execution.

Defects4J~\cite{just2014defects4j}, Bugs.jar~\cite{saha2018bugs}, Bears~\cite{madeiral2019bears} and BugSwarm~\cite{tomassi2019bugswarm} contain executable bugs from real-world repositories.
However, Bugs.jar and Bears face serious execution and reproducibility challenges due to missing dependencies and incomplete configuration environments~\cite{zhu2023reproducibility}.
Also, Durieux and Abreu state that 96.4\% of the BugSwarm benchmark is not suitable for APR and FL, for reasons that include duplicate samples, lack of failing tests, and changes to non-source code files~\cite{durieux2019critical}.

Finally, Defects4J is a milestone of benchmark research, containing bugs that are reproducible and adequate for APR and FL.
Yet, Defects4J mostly contains old bugs, at the time of writing, it was last updated in 2020.
Indeed, Silva et al.~\cite{silva2021flacoco} find that the majority of Defects4J bugs require Java 6 or earlier bytecode while Java 6 is no longer supported by Oracle as of December 2018.
Moreover, given that the cutoff date of most training datasets used for LLMs is beyond 2020, there exists a significant risk that Defects4J examples are included in them, thus threatening the validity of recent evaluations of LLM-based techniques on Defects4J.

To the best of our knowledge, \name is the first tool for collecting executable bug-fix benchmarks that are sourced from the real world and fully reproducible, appropriate for building new benchmarks containing recent bug-fixes.

ActionsRemaker~\cite{zhu2023actionsremaker} is a tool for reproducing GitHub Actions builds.
We favor using \textit{Act} instead of ActionsRemaker 
because \textit{Act} is a popular and mature open-source tool, with significantly higher reliability compared to academic prototyping.
To the best of our knowledge, we are the first to leverage GitHub Actions to collect bug-fix commits.

%% file: conference_101719.bbl
\begin{thebibliography}{10}
\providecommand{\url}[1]{#1}
\csname url@samestyle\endcsname
\providecommand{\newblock}{\relax}
\providecommand{\bibinfo}[2]{#2}
\providecommand{\BIBentrySTDinterwordspacing}{\spaceskip=0pt\relax}
\providecommand{\BIBentryALTinterwordstretchfactor}{4}
\providecommand{\BIBentryALTinterwordspacing}{\spaceskip=\fontdimen2\font plus
\BIBentryALTinterwordstretchfactor\fontdimen3\font minus
  \fontdimen4\font\relax}
\providecommand{\BIBforeignlanguage}[2]{{%
\expandafter\ifx\csname l@#1\endcsname\relax
\typeout{** WARNING: IEEEtran.bst: No hyphenation pattern has been}%
\typeout{** loaded for the language `#1'. Using the pattern for}%
\typeout{** the default language instead.}%
\else
\language=\csname l@#1\endcsname
\fi
#2}}
\providecommand{\BIBdecl}{\relax}
\BIBdecl

\bibitem{sim2003using}
S.~E. Sim, S.~Easterbrook, and R.~C. Holt, ``Using benchmarking to advance
  research: A challenge to software engineering,'' in \emph{25th International
  Conference on Software Engineering, 2003. Proceedings.}\hskip 1em plus 0.5em
  minus 0.4em\relax IEEE, 2003, pp. 74--83.

\bibitem{wright2010validity}
H.~K. Wright, M.~Kim, and D.~E. Perry, ``Validity concerns in software
  engineering research,'' in \emph{Proceedings of the FSE/SDP workshop on
  Future of software engineering research}, 2010, pp. 411--414.

\bibitem{just2014defects4j}
R.~Just, D.~Jalali, and M.~D. Ernst, ``Defects4j: A database of existing faults
  to enable controlled testing studies for java programs,'' in
  \emph{Proceedings of the 2014 international symposium on software testing and
  analysis}, 2014, pp. 437--440.

\bibitem{jacovi2023stop}
A.~Jacovi, A.~Caciularu, O.~Goldman, and Y.~Goldberg, ``Stop uploading test
  data in plain text: Practical strategies for mitigating data contamination by
  evaluation benchmarks,'' \emph{arXiv preprint arXiv:2305.10160}, 2023.

\bibitem{zhang2023critical}
Q.~Zhang, T.~Zhang, J.~Zhai, C.~Fang, B.~Yu, W.~Sun, and Z.~Chen, ``A critical
  review of large language model on software engineering: An example from
  chatgpt and automated program repair,'' 2023.

\bibitem{zhu2023reproducibility}
H.-N. Zhu and C.~Rubio-Gonz{\'a}lez, ``On the reproducibility of software
  defect datasets,'' \emph{ICSE. IEEE}, 2023.

\bibitem{tomassi2019bugswarm}
D.~A. Tomassi, N.~Dmeiri, Y.~Wang, A.~Bhowmick, Y.-C. Liu, P.~T. Devanbu,
  B.~Vasilescu, and C.~Rubio-Gonz{\'a}lez, ``Bugswarm: Mining and continuously
  growing a dataset of reproducible failures and fixes,'' in \emph{2019
  IEEE/ACM 41st International Conference on Software Engineering (ICSE)}.\hskip
  1em plus 0.5em minus 0.4em\relax IEEE, 2019, pp. 339--349.

\bibitem{madeiral2019bears}
F.~Madeiral, S.~Urli, M.~Maia, and M.~Monperrus, ``Bears: An extensible java
  bug benchmark for automatic program repair studies,'' in \emph{2019 IEEE 26th
  International Conference on Software Analysis, Evolution and Reengineering
  (SANER)}.\hskip 1em plus 0.5em minus 0.4em\relax IEEE, 2019, pp. 468--478.

\bibitem{JetBrainsCI}
``The state of developer ecosystem 2022,''
  \url{https://www.jetbrains.com/lp/devecosystem-2022/team-tools/#ci-tools},
  accessed: 2023-09-29.

\bibitem{lin2017quixbugs}
D.~Lin, J.~Koppel, A.~Chen, and A.~Solar-Lezama, ``Quixbugs: A multi-lingual
  program repair benchmark set based on the quixey challenge,'' in
  \emph{Proceedings Companion of the 2017 ACM SIGPLAN international conference
  on systems, programming, languages, and applications: software for humanity},
  2017, pp. 55--56.

\bibitem{tan2017codeflaws}
S.~H. Tan, J.~Yi, S.~Mechtaev, A.~Roychoudhury \emph{et~al.}, ``Codeflaws: a
  programming competition benchmark for evaluating automated program repair
  tools,'' in \emph{2017 IEEE/ACM 39th International Conference on Software
  Engineering Companion (ICSE-C)}.\hskip 1em plus 0.5em minus 0.4em\relax IEEE,
  2017, pp. 180--182.

\bibitem{majd2019code4bench}
A.~Majd, M.~Vahidi-Asl, A.~Khalilian, A.~Baraani-Dastjerdi, and B.~Zamani,
  ``Code4bench: A multidimensional benchmark of codeforces data for different
  program analysis techniques,'' \emph{Journal of Computer Languages}, vol.~53,
  pp. 38--52, 2019.

\bibitem{prenner2023runbugrun}
J.~A. Prenner and R.~Robbes, ``Runbugrun--an executable dataset for automated
  program repair,'' \emph{arXiv preprint arXiv:2304.01102}, 2023.

\bibitem{jiang2023impact}
\BIBentryALTinterwordspacing
N.~Jiang, K.~Liu, T.~Lutellier, and L.~Tan, ``Impact of code language models on
  automated program repair,'' in \emph{Proceedings of the 45th International
  Conference on Software Engineering}, ser. ICSE '23.\hskip 1em plus 0.5em
  minus 0.4em\relax IEEE Press, 2023, p. 1430–1442. [Online]. Available:
  \url{https://doi.org/10.1109/ICSE48619.2023.00125}
\BIBentrySTDinterwordspacing

\bibitem{csuvik2022fixjs}
V.~Csuvik and L.~Vid{\'a}cs, ``Fixjs: a dataset of bug-fixing javascript
  commits,'' in \emph{Proceedings of the 19th International Conference on
  Mining Software Repositories}, 2022, pp. 712--716.

\bibitem{avulaminecraft}
S.~K. Avula, V.~Vobbilisetti, and S.~Mondal, ``Minecraft: Automated mining of
  software bug fixes with precise code context,'' in \emph{Proceedings of the
  38th IEEE/ACM International Conference on Automated Software Engineering},
  2023.

\bibitem{saha2018bugs}
R.~K. Saha, Y.~Lyu, W.~Lam, H.~Yoshida, and M.~R. Prasad, ``Bugs. jar: A
  large-scale, diverse dataset of real-world java bugs,'' in \emph{Proceedings
  of the 15th international conference on mining software repositories}, 2018,
  pp. 10--13.

\bibitem{durieux2019critical}
T.~Durieux and R.~Abreu, ``Critical review of bugswarm for fault localization
  and program repair,'' \emph{arXiv preprint arXiv:1905.09375}, 2019.

\bibitem{silva2021flacoco}
A.~Silva, M.~Martinez, B.~Danglot, D.~Ginelli, and M.~Monperrus, ``Flacoco:
  Fault localization for java based on industry-grade coverage,'' \emph{arXiv
  preprint arXiv:2111.12513}, 2021.

\bibitem{zhu2023actionsremaker}
H.-N. Zhu, K.~Z. Guan, R.~M. Furth, and C.~Rubio-Gonz{\'a}lez,
  ``Actionsremaker: Reproducing github actions,'' \emph{ICSE-Companion. IEEE},
  2023.

\end{thebibliography}
